Ataklti Kahsu
Department of Information System
Adama Science and Technology University
Adama, Ethiopia
atuka2@gmail.com

Solomon Teferra (PhD)
School of Information Science
Addis Ababa University
Addis Ababa, Ethiopia
solomon_teferra_7@yahoo.com

# Abstract

This thesis proposes and describes a research attempt at designing and developing a speaker-independent spontaneous automatic speech recognition system for Tigrigna. The acoustic model of the Speech Recognition System is developed using Carnegie Mellon University's Automatic Speech Recognition development tool (Sphinx), while the SRIM tool is used for the development of the language model.

# I. INTRODUCTION

Information processing machines have become ubiquitous. However, the current modes of human-machine communication are geared more towards living with the limitations of computer input and output devices than the convenience of humans. Speech is the primary mode of communication among human beings. On the other hand, the most prevalent means of input to computers is through a keyboard or a mouse. It would be nice if computers could listen to human speech and carry out their commands [1].

Speech recognition is a field of computer science that deals with designing computer systems that recognize spoken words. It is a technology that allows a computer to identify the words that a person speaks through a microphone or telephone. Speech recognition can be defined as the process of converting an acoustic signal captured by a microphone or a telephone into a set of words [2] [3]. ASR is one of the fastest-developing fields in the framework of speech science and engineering. As the new generation of computing technology, it comes as the next major innovation in man-machine interaction, after the functionality of TTS and supporting IVR systems.

The first attempts (during the 1950s) to develop techniques in ASR, which were based on the direct conversion of speech signals into a sequence of phoneme-like units, failed [4]. The first positive results of spoken word recognition came into existence in the 1970s, when general pattern matching techniques were introduced. As the extension of their applications was limited, the statistical approach to ASR started to be researched during the same period. Nowadays, statistical techniques prevail over ASR applications.

Common speech recognition systems these days can recognize thousands of words. The last decade has witnessed dramatic improvements in speech recognition technology, to the extent

Large Vocabulary Spontaneous Speech Recognition for Tigrigna: high-performance algorithms and systems are becoming available. In some cases, the transition from laboratory demonstration to commercial deployment has already begun [3]. The reason for the evolution of ASR, hence its improvement, is due to its applications in many aspects of our command recognition, information inquiry, dictation, personal computer interfaces, automated telephone services, interactive voice response, and special-purpose commercial and industrial systems. It can also be used in education for language learning. It has other important applications for handicapped people, helping them with their daily lives and their communication with the rest of society.

## II. Related Works

A lot of work has been done on automatic speech recognition systems for different languages. These works range from activities involving the development of isolated speech recognition systems to the development of spontaneous speech recognition systems. The main area on which this is focused is to develop a spontaneous ASR system for Tigrigna. Therefore, conducting research related to the subject area of speech recognition has a vital role in choosing appropriate mechanisms for solving problems and enhancing the performance of the recognizer.

## Speech recognition for other languages

Different kinds of research have been conducted for the past decade in the area of automatic speech recognition, to mention the following: Speaker-independent Dictation of Chinese Speech with 32K Vocabulary: they developed speaker-independent, word-based dictation. A deliberately designed 120-speaker database was built for training; inter-syllable context, tonal, and endpoint-dependent acoustic models are applied with promising MFCC features; Two-pass acoustic matching accelerates recognition, taking full advantage of the monosyllabic structure of Chinese speech. A complete word bigram and trigram serve as language processing modules. They reported that the system reaches 90% character accuracy, performing in almost real-time on a Pentium PC [21].

Large Vocabulary Continuous Speech Recognition in Greek: Corpus and an Automatic Dictation System: in this work the researchers presents the creation of the first Greek Speech Corpus and the implementation of a Dictation System for workflow improvement in the field of journalism. They evaluated the word error rate (WER) and the recognition time (xRT) for the three Acoustic Models: Unisex for the gender-independent model they have got 21.01% 6.16 xRT for Male gender dependent model they have got 19.27% WER and 5.78xRT for Female gender Dependent model they have got 20.85 WER and 5.85xRT. As the final step of their work, they developed a graphical user interface (GUI) in order to complete the Greek Dictation System, named Logotypografos1.0. This tool was created using the MFC C++ library for the graphical interface. Automatic Speech Recognition for Amharic also started in 2001 when [22] developed a speech recognizer for Amharic that recognizes a sub-set of isolated consonant-vowel (CV) syllable using the HTK (Hidden-Markov Modeling Toolkit). Forty one CV syllables of Amharic language out of 234 were selected. Speech data of the selected CV syllables has been recorded from eight people (4 male and 4 female) with the age range of 20 to 33 years. The average recognition accuracies achieved were 87.68% and 72.75% for speaker dependent and speaker independent systems, respectively. As indicated by the researcher, the result seems to be low compared to systems developed for other languages. This might be due to problems of the recording environment and insufficient training speech data.

As a continuation of [22] [14] had developed sub-word based isolated word recognition systems for Amharic using HTK. The sub-word units used in his experiment are phones, tri-phones, and CV-syllables. He considered 20 phones (out of 39) and 104 CV syllables, which are formed using the selected phones. Speech data of 170 words, which are composed of the selected sub-word units, have been recorded from 20 speakers (speech of 15 speakers for training and the remaining for testing). Speaker dependent phone-based and tri-phone-based systems have an average recognition accuracy of 83.07% and 78% respectively.

Furthermore [9] and [10] in 2015 have emerged as continuation of the development of Amharic speech recognition by different researchers which leads from read speech to Spontaneous Speech Recognition System. [9] Developed Spontaneous speech recognizer for Amharic, using HTK, with speech data size of 36 speakers with 3 hours and 10 minutes of training data. The test data was prepared with 104 utterances consisting of 820 unique words from 14 speakers, which is 10 of them involved in training and 4 of them did not involve in training. The Language Model used was bigram Language Model developed using HLStats and word networks built from this bigram using HBuild tool.

Another important step towards the development of Amharic speech recognition also Developed and implemented two different dictation application systems by [10].The first is dictation application of BRANA (ብራና) in which the entire application is developed with java programming language and it is a PPT (push to talk) prototype which shows a good performance in writing down the recognized results as per the speech recognizer performance.

According to [10] the system performs 50% word accuracy with small training corpus (90 min) and the performance also increases as the training corpus increases. In this work testing of the application with continuous (read) speech recognizer was also performed. The experimental results show that performance of continuous speech recognizer is better than that of the spontaneous speech recognizer that shows the system to perform better on the recognizer that have a better recognition accuracy.

## *Speech recognition for Tigrigna language*

Research in Automatic Speech Recognition for Tigrigna is not much matured as that of Amharic. The only (as the knowledge of the researcher) available is [7] which is entailed as HMM Based Large Vocabulary, Speaker Independent Continuous Tigrigna Speech Recognition conducted in HTK environment.

The researcher, in his work, used database comprised of 250 utterances that are used for training and 50 sentences for testing and evaluation. The data is pre-processed in line with the requirements of the HTK toolkit. Furthermore the researcher attempted to build speech to text conversion for Tigrigna language using the statistical approach. In order to support the Acoustic Models, a bigram Language Model is constructed. In addition, pronunciation dictionary is prepared and used as an input. In addition to the mono-phone based speech recognizer, the researcher tries a tri-phone based system in which the left and right contexts consider and modeled. According to the researcher, performance tests also conducted at various stages using the training and test data. As a result of this the researcher arrived at 60.20% word level correctness, 58.97% word accuracy, and 20.06% sentence level correctness are obtained.

# III. EXPERIMENTATION

This chapter describes the design and construction of a Spontaneous speech recognizer for Tigrigna language. First the requirements for the system data preprocessed are stated and the development platform and tools are described. Next an outline of the steps needed to build a speech recognizer is given. Subsequent sections explain these steps in more detail.

5.1 Introduction

In this study an attempt has been made to design and construct spontaneous speech recognizer for Tigrigna language that is capable of recognizing Tigrigna speech. In this attempt, to increase model resolution several experiments have been performed investigating different possibilities of improvement. When building statistical models, the possibility of high performance is always based on a trade of between model complexity and the amount of available data for parameter estimation. A good choice for model structure, based on the comprehension of the sources of variability in the physical phenomena, can help increasing the ability of the model to fit the problem of interest with the minimum number of parameters.

The intention was developing a system that is as general as possible, so that it can be used, with slight modifications and some adaptive training, as a speech interface for many other different applications.

To meet these requirements the recognizer was designed to recognize large vocabulary, spontaneous speech data. This was implemented using phonemes as base unit. Thus, the systems vocabulary was not limited in any way to some fixed set of words. If a new vocabulary is required to be incorporated, what needs to be done is only modifying the Language Model and adding the new word to the pronunciation dictionary. This means that the system can easily be extended and adapted to many applications.

During this study two main directions of improvement have been followed. Context independent (CI) model as a first method. And the second method context dependent (CD) models have been tested with the same data which is a model to add Gaussian terms in the state output probability distribution. In this case the re-estimation procedure is completely responsible for adapting model parameters to different sources of variability in the speech signal.

## IV. The Data Set

The trainer learns the parameters of the models of the sound units using a set of sample speech signals. This is called a training database. The database contains information required to extract statistics from the speech the form of the Acoustic Model. Database should have enough speakers recording, variety of recording conditions, enough acoustic variations and all possible linguistic sentences [17] the construction of such database is a major undertaking by itself and involves a number of subtasks. It also demands sufficiently large time and big financial support.

For steps in sub-section 5.2.1 a corpus containing 3524 sentences spoken by twenty four speakers of twelve male and twelve female were prepared as in Table 2. The number of female speakers and male speakers are balanced. The utterances are also comprised of all Tigrigna phonemes in many phonetic contexts. Due to this this system may not work for non-native speakers.

A data set must have two parts, training part which is used for training the system and test part used for testing the system's performance Most of the data from the database, consisting of ten female with 1617 sentences and male with 1558 sentences, are used. Usually test part is about 1/10th of the full data size [17]. The remaining two males with a total number of 172 sentences and two females with a total number of 177 sentences were used for test part.

**Feature vector extraction: -** Another important task in data preparation is the extraction of feature vectors i.e. parameterizing the raw speech waveforms into sequences of feature vectors. For training and testing purposes it is reasonable to do this once for all and save the feature vectors in a file called feat. Because each utterance is processed a number of times during training it helps to reduce later processing costs. The Sphinx trainer is responsible for translating audio files to feature vector files using Mel scale cepstral coefficients with input parameters listed in table 1.

*Table 1:- Parameters used in feature extraction*

| Parameter | Value |
|---|---|
| Sampling Rate | 16Khz |
| Pre-emphases Coefficient | 0.97 |
| Window Size | 25ms |
| Overlap Duration | 15ms |
| Hamming Window | True |
| Zero Mean | True |
| Cepstral lifter | 22 |
| Number of Cepstral Coefficient | 12 |

# V.  Language Model

After training, it's mandatory to run the decoder to check training results. The Decoder takes a model, tests part of the database and references transcriptions and estimates the quality of the model. During the testing stage, Language Model with the description of the order of words in the language is crucial part of a Speech Recognition System. The Language Model describes the probabilities of the sequences of words in the text and is required for speech recognition there is no Language Model prepared for Tigrigna language yet. In order to evaluate the decoder, Language Model must be prepared first. Therefore, in order to build language model the following steps were followed.

# VI.  Data collection

The first step of Language Model building is a collection of the data. The amount of data needed depends on the domain and vocabulary. Usually for a good model it needs a significant amount of texts - at least 100mb [20]. This data needs to get text by transcribing existing recordings, collecting data from the web, generating it artificially with scripts and from broadcasting. Generally, the most valuable data is a real-life data. But this task by itself requires sufficiently large time. For this work data were collected form Woyan gazette, Mekalih Tigray gazette and Tigray television an amount of 50MB.

**Text cleanup: -** Once data is collected it must be cleaned - punctuation removed, sentences split, numbers expanded to text representation. For this purpose a system was used splitting paragraphs in to sentence level and it enables representation of numbers to their expanded text.

**Language Model Preparation:-**The Unigram, Bigram or Trigram Language Models can be generated using either the CMU Language Modeling CMU-LM toolkit or SRILM toolkit which works in Linux environment. During language model preparation the SRILM language modelling tool was used.

**SRILM:** SRILM is a toolkit for building and applying statistical Language Models (LMs), primarily for use in speech recognition. Training with SRILM is easy, and recommend by [17]. Moreover, SRILM is the most advanced toolkit and up to date. The SRILM toolkit also useful as it allows more options for generating the Language Models such as different smoothing techniques like Laplace smoothing, Additive smoothing and linear interpolation. For preparing the Language Model 10MB normalized text with 30937 sentences was used. In order to optimize the language model the Laplace smoothing technique was used during the development. After generating, the Language Model has to be converted to binary file using a converter. Sphinx4 requires DMP format. DMP format can produce file with Sphinxss_lm_convert command from Sphinx base. Finally data required for training and testing are organized as follow:-

    TigSpeech.dic - Phonetic dictionary

    TigSpeech.phone - Phoneset file

    TigSpeech.lm.DMP - Language Model

    TigSpeech.filler - List of fillers

    TigSpeech_train.fileids - List of files for training

    TigSpeech_train.transcription - Transcription for training

    TigSpeech_test.fileids - List of files for testing

    TigSpeech_test.transcription - Transcription for testing

**Setting up configuration files:-**Before starting the train parameters needed to edit the configuration files in "etc" folde.

Setup the format of database audio:-As stated the audio files used for training the model are in .WAV format before starting train this needs to be configured as follow

$CFG_WAVFILES_DIR = "$CFG_BASE_DIR/wav";

$CFG_WAVFILE_EXTENSION = 'wav';

$CFG_WAVFILE_TYPE = 'mswav';

**Configure model type and model parameters:-**CMU Sphinx supports different types of Acoustic Models; continuous, semi- continuous and phonetically tied. In this work, the training was performed on the continuous (.cont.) HMM type with Context-Independent (CI) and context-dependent (CD) phones model. The Acoustic Model is developed with no skip transition topology and 8 Gaussian mixtures as follows.

$CFG_HMM_TYPE = '.cont.' #

#$CFG_HMM_TYPE = '.semi.' #

#$CFG_HMM_TYPE = '.ptm.' #

**Configure sound feature parameters:-**Sphinx uses a rate of 16 thousand samples per second (16 KHz) with 16 bit sample size then this has to be configured in alignment with the audio format and sample rate

# Feature extraction parameters

$CFG_WAVFILE_SRATE = 16000.0;

Table 2: Dataset used

| Speaker Code | Gender | Number of Sentences | Number of non-speech events |
|---|---|---|---|
| spF21 | Female | 96 | 22 |
| spF23 | Female | 81 | 19 |
| spM004 | Male | 69 | 75 |
| spM007 | Male | 103 | 80 |

# VII. Results with modeling non-speech

The training was performed in two phases based on non-speech events with CI and CD and the same acoustic and Language Model containing four unseen speaker (speaker whom are not participated in the training) with a total of 349 sentences were used. Table 3 shows speakers used during testing with their gender, number of sentences and amount of modeled non speech events. On his step non speech events (such as speaker generated, other speaker generated and back ground noise) were modeled. In training transcription 1285 Filled pause (++FP++), 19 Breaths (++BR++), 228 Hesitations (++HES++), 10 Lip smacks (++LP++) and one through clear (++THC++) with respect to speaker generated events. And also 70 other speakers (++OTH++) and 26 back ground noises were modeled. In testing transcription 170 Filled pause (++FP++), one Breaths (++BR++), 12 Hesitations (++HES++), one Lip smacks (++LP++) and zero through clear (++THC++) with respect to speaker generated events. And one other speaker (++OTH++) and two back ground noise were also modeled.

*Table 3*: Results of CD Model with eight Gaussian mixture

| Speaker Code | % Correct | % Accuracy | % SER |
| --- | --- | --- | --- |
| spF21 | 32.97% | 30.13% | 85.50% |
| spF23 | 27.36% | 37.01% | 100.00% |
| spM004 | 37.09% | 36.36% | 83.51% |
| spM007 | 35.97% | 35.37% | 100.00% |
| Total | 36.83% | 36.83% | 86.37% |

Table 3 shows recognition result of CD Acoustic Model with eight (8) Gaussian Mixture and with overall recognition accuracy 36.83%. On this stage highest performance improvement is achieved from all the Gaussian mixtures Model. In this model the word level accuracy is increased. As shown from the results the eight (8) Gaussian Mixture performance is accurate than the four (4) Gaussian Mixture by 3.50% and 6.71% from two (2) Gaussian Mixture and finally 7.28% from the single Gaussian Mixture. Form the results obtained, As Gaussian mixture increases the overall recognition accuracy increase. But the sentence level accuracy is very low in all Gaussian mixtures.

*Table 4:* Summarized Results of all Gaussian mixture CD Model

| Gaussian | % Correct | % Accuracy | % SER |
| --- | --- | --- | --- |

| mixture | | | |
|---|---|---|---|
| Single | 30.57% | 29.55% | 86.98% |
| Two (2) | 31.12% | 30.12% | 86.41% |
| Four (4) | 33.33% | 33.33% | 86.67% |
| Eight (8) | 36.83% | 36.83% | 86.37% |

## VIII. Architecture of Developed Speech Recognizer

Figure 1 shows the components of developed speech recognizer as it processes a single utterance taken form the data set indicating the computation of the prior and likelihood. In the figure shows the recognition process in three stages. In the feature extraction or signal processing stage, the acoustic waveform is sampled into frames which are transformed into spectral features. Each time window is thus represented by a vector of around 39 features representing this spectral information as shown in table 4

In the acoustic modeling or phone recognition stage, the system compute the likelihood of the observed spectral feature vectors given linguistic units the output of this stage is as a sequence of probability vectors, one for each time frame, each vector at each time frame containing the likelihoods that each phone unit generated the acoustic feature vector observation at that time.
Finally, in the decoding phase, the acoustic model (AM), which consists of the sequence of acoustic likelihoods, plus the dictionary of word pronunciations (Tigrigna Dictionary), combined with the language model (LM) (Trigram Tigrigna Language Model), and output the most likely sequence of words.

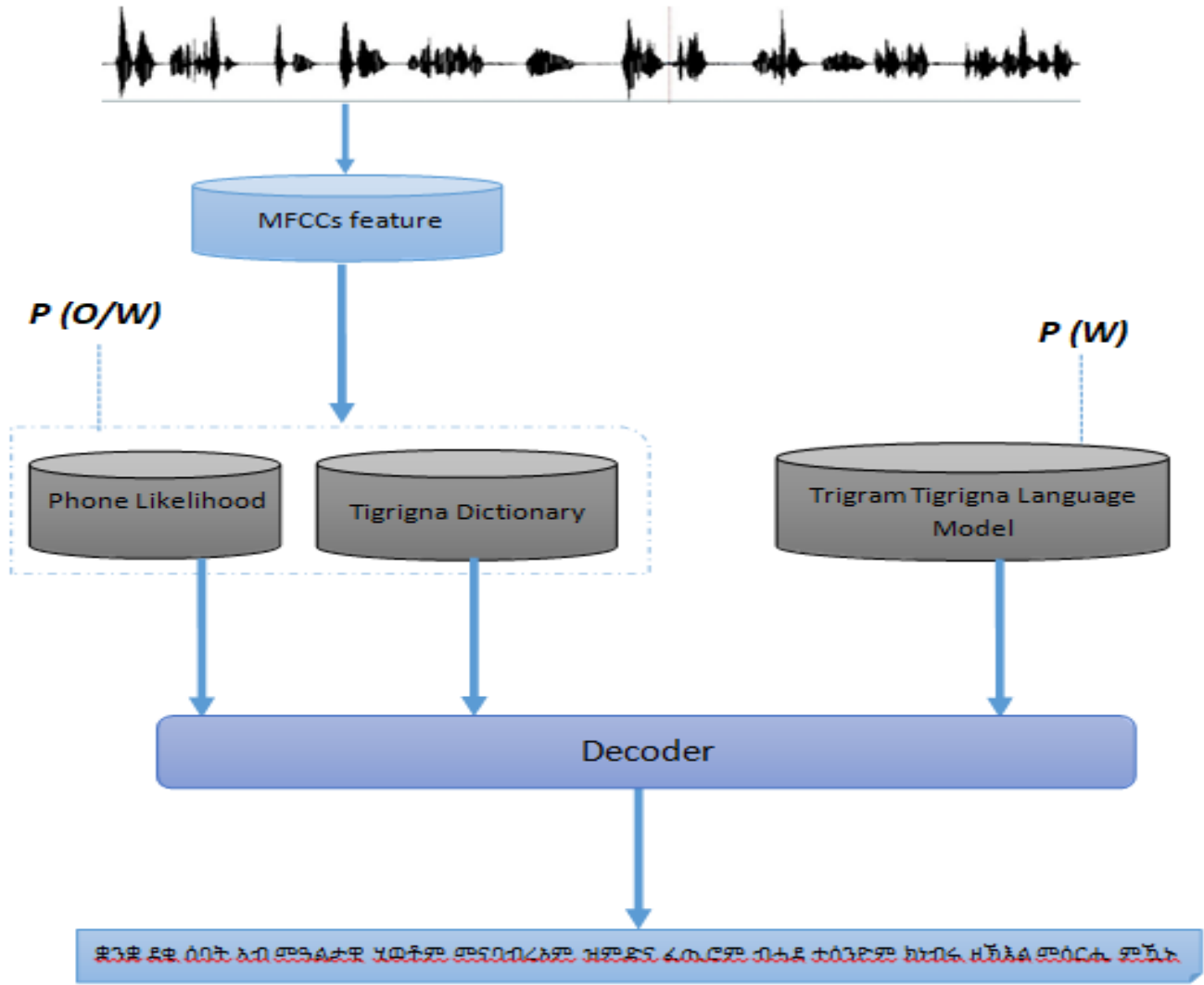

Figure 1: Architecture of developed speech recognizer decoding a single sentence

## IX. Recommendations

From the experimental results and from the point view of this study, an attempt was made to focus on investigating the possibility of developing, large vocabulary spontaneous speech recognition for Tigrigna. Depending on finings and challenges facing during the development followings are the possibilities that should have to be done for the future in order to boost the performance of the recognizer.

As stated in section 1.7.2 there is no spontaneous speech corpus prepared previously, therefore in this work an attempt has made on developing Spontaneous preparing the required data set. But it does not mean that the data set is sufficient therefore in the future researchers whom work on this area have to increasing the size of the data set, doing this helps on improving the recognizer accuracy and even to ensure the applicability of ASR. Since it is difficult to use ASR with low accuracy in a real world

In this study the experiment was conducted with using tri-gram language model with Laplace smoothing techniques in order to remove zero probabilities and only the one with less Perplexity was selected. For future the language model has to be increased in size as well as the system has to be experimented with different smoothing techniques and increasing tri-gram to other N-grams will improve the recognizer accuracy.

One of the difficult in spontaneous speech is non-speech events (dis-fluencies), therefore handling none speech events has a big role in recognizer's performance. In this work an attempt has been made on handling these issues, but still it is possible to handle them further by applying different techniques. One of the task one can do in order to handle these non-speech events effect is including or removing from acoustic, lexical and language model, by studying their nature in detail and separately.

In this research a recognized (transcribed) pronunciation dictionary was used, but there are words with the same meaning but uttered differently by different speakers (such as □□□ (two) and □□□ (two)), dialect and accent variations were not considered in this work. Such variations decrease the performance of the recognizer. Therefore it is better to use alternative pronunciation dictionary in order to improve recognizer performance.

As indicated on this work database used was not restricted in any domain and the data was collected from different sources whereas [10] develops dictation system for Amharic language and founds good result so it is valuable to develop domain dependent system for Tigrigna.